\documentclass[doublecol]{epl2}
\usepackage{graphicx}
\usepackage{amsmath}
\usepackage{bm}
\usepackage{verbatim}
\usepackage{color}
\usepackage{ulem}
\DeclareMathOperator{\sgn}{sgn}
\renewcommand\footnotemark{}
\def\be{\begin{equation}}
\def\ee{\end{equation}}

\def\bi{\begin{itemize}}
\def\ei{\end{itemize}}
\def\bn{\begin{enumerate}}
\def\en{\end{enumerate}}
\def\bea{\begin{eqnarray}}
\def\eea{\end{eqnarray}}
\def\no{\nonumber}
\def\ba{\begin{array}}
\def\ea{\end{array}}
\def\bd{\begin{displaymath}}
\def\ed{\end{displaymath}}

\title{Quantum-classical equivalence and ground-state factorization}

\author{Jahanfar Abouie \inst{1,3}  \thanks{E-mail: \email{jahan@iasbs.ac.ir}} \and Reza Sepehrinia \inst{2,3} \thanks{E-mail: \email{sepehrinia@ut.ac.ir}}\thanks{The authors have contributed equally to this work.}}

\shortauthor{J. Abouie and R. Sepehrinia}

\institute{
\inst{1} Department of Physics, Institute for Advanced Studies
in Basic Sciences (IASBS), Zanjan 45137-66731, Iran\\
\inst{2} Department of Physics, University of Tehran, Tehran 14395-547,
Iran\\
\inst{3}  School of Physics, Institute for Research in
Fundamental Sciences, IPM, Tehran 19395-5531, Iran\\
}


\abstract{
We have performed an analytical study of quantum-classical equivalence for quantum $XY$-spin chains with arbitrary interactions to explore the classical counterpart of the factorizing magnetic fields that drive the system into a separable ground state.
We demonstrate that the factorizing line in parameter space of a quantum model is equivalent to the so-called natural boundary that emerges in mapping the quantum $XY$-model onto the two dimensional classical Ising model.
As a result, we show that the quantum systems with the non-factorizable ground state could not be mapped onto the classical Ising model. Based on the presented correspondence we suggest a promising method for obtaining the factorizing field of quantum systems through the commutation of the quantum Hamiltonian and the transfer matrix of the classical model.}

\begin{document}

\maketitle
\section{Introduction}\label{Intro}
One of the essential challenges in condensed matter physics is that of
solving many body interacting systems. One major class of these systems are
quantum spin models which despite their simplicity capture various complex
physical phenomena. Even among these idealized models we rarely and only in
low dimensions encounter solvable cases. Therefore in many systems it is very
valuable to find even a single eigenstate of the Hamiltonian. It is found
that in general non-exactly solvable models admit an exact \textit{factorized}
ground state for special values of Hamiltonian parameters e.g. external
transverse magnetic field \cite{kurmann1982antiferromagnetic, giampaolo2008theory,  giampaolo2009separability, rossignoli2009factorization, rezai2010factorized, abouie2012ground}. The analysis of quantum entanglement contained
in the ground state has revealed additional features of the factorization point.
Unlike the standard magnetic observable the entanglement displays an anomalous
behavior at the factorizing field (in addition to the \textit{critical} field) and vanishes at this point.
Across this point the system undergoes an entanglement phase transition \cite{amico2009entanglement,roscilde2004studying}.
Furthermore the quantum discord, another measure of quantum correlations,
 exhibits scaling behavior close to factorization point \cite{ciliberti2010quantum,tomasello2011ground,campbell2013criticality,sarandy2013quantum} which is
of collective nature but different from a quantum phase transition which is
accompanied by a change of symmetry. For finite systems it is also
shown that the ground state remains entangled as the factorizing field is approached
and undergoes parity transition across this point \cite{rossignoli2008entanglement,canosa2010separability,giorgi2009ground}.

A fairly general approach which allows us to make precise statements about
ground state of quantum systems is mapping the $d$-dimensional quantum system
onto a $d+1$-dimensional classical system. Then the different properties of quantum model
like orders, correlations and response functions and scaling behavior near quantum
critical points can be studied through the relation to their classical counterparts.
However, it turns out that mapping onto an specific classical model holds
only for a restricted region of parameter space of quantum model.
For instance the ground state of the spin-$\frac{1}{2}$ quantum $XY$-chain with uniform interactions in the presence of a magnetic field is equivalent to
a two-dimensional (2D) rectangular Ising model but this equivalence is
restricted to the region outside a circle which is called \textit{natural boundary} \cite{suzuki1971relationship}. This circular boundary
has been also observed in studying rather different aspects of the $XY$-model. One of them is the spin-spin correlations of the quantum $XY$-chain which exhibit non-oscillatory (classical) asymptotic behavior outside this circle whereas
they have oscillatory (quantum) asymptotic behavior inside it \cite{barouch1971statistical}. Another property of this circle was found in investigation of the ground state entanglement of the quantum $XY$-model. Along this circle the entanglement vanishes in correspondence with an exactly
factorized state \cite{amico2009entanglement} and across it an entanglement phase transition from odd to even parity ground state occurs\cite{roscilde2004studying}.

In this letter, motivated by these observations, we perform a detailed investigation to find out the relationship between the factorization of the ground state of quantum $XY$-chains with arbitrary interactions and their mappability onto the classical Ising model. We show that the above mentioned properties are not restricted to the particular case of uniform couplings and
are maintained for the general $XY$-chains with arbitrary interactions.
We obtain the natural boundary for the general case and show that this boundary coincides with
the factorizing line of the quantum chain. We apply our results to several examples which
do or do not have factorizing field.
We find that there is no equivalent Ising model for the quantum models
which do not possess a factorized ground state.

\section{General XY model and ground state factorization}\label{Fac}

The Hamiltonian of a spin-$\frac12$ quantum $XY$-chain in the presence of a magnetic field can be written as:
\be
\label{m.b.h}
\mathcal{H}_q=-\sum_{i,r}
(J^x_{i,i+r} \sigma^x_{i}\sigma^x_{i+r}+J^y_{i,i+r}\sigma^y_{i}\sigma^y_{i+r})-\sum_{i}h_i\sigma^z_i,
\ee
where $\sigma^{x,y}$ are Pauli matrices and $h_i$ is a general transverse magnetic field.
The interaction between two spins depends on the position $i$ and the distance $r$ via $J^{\mu}_{i,i+r}$ where $\mu=x,y$
and are assumed to be ferromagnetic ($J^{\mu}_{i,i+r}>0$) or antiferromagnetic ($J^{\mu}_{i,i+r}<0$) such that there is no frustration in the system. The above Hamiltonian possesses a factorized state as $|FS\rangle=\otimes_i|\theta_i\rangle$ at the factorizing fields \cite{kurmann1982antiferromagnetic,rossignoli2009factorization};
\bea
\label{cond}h_i^f &=& \cot \theta_i \sum_{r}(J^x_{i,i+r} \sin \theta_{i+r}+J^x_{i,i-r} \sin \theta_{i-r}),
\eea
where $|\theta_i\rangle=\cos\frac{\theta_i}{2}|+\rangle_i+\sin\frac{\theta_i}{2}|-\rangle_i$,
and $\sigma_i^z|\pm\rangle_i=\pm|\pm\rangle_i$.
Here,  $\theta_i\in(-\pi/2,\pi/2)$ is the acute angle of the spin $\sigma_i$ with $+z$ axis given by relation
\be \label{coscos}
\cos \theta_i \cos \theta_{i+ r}=J^y_{i,i+ r}/J^x_{i,i+ r}.
\ee
The equation (\ref{coscos}) requires $|J^y_{i,i+ r}/J^x_{i,i+ r}|\leq 1$ which ensures that the factorized state $|FS \rangle$ is the ground state \cite{rossignoli2009factorization}.
For chains with nearest-neighbor interactions, $J^{x,y}_{i,i+r}=J^{x,y}_i \delta_{r1}$, the factorizing field (\ref{cond}) is simplified to
\bea
\label{ff}
 h_i^f =  \cot \alpha_i(\sin \alpha_{i+1} |J^x_{i}|   + \sin \alpha_{i-1} |J^x_{i-1}|),
\eea
where we have defined $\theta_{i}=\xi_i\alpha_{i}$, with $\alpha_i\in(0,\pi/2)$ and $\xi_i$ is the sign of $\theta_{i}$ satisfying $\xi_{i}=\text{sgn} J_{i-1}^x \xi_{i-1}$ .
The Hamiltonian (\ref{m.b.h}) has the global $z$-parity symmetry and the factorized ground state
has a degenerate partner state, $|-FS\rangle=\otimes_i|-\theta_i\rangle$.

\begin{figure}[t]
\centerline{\includegraphics[width=8.5cm]{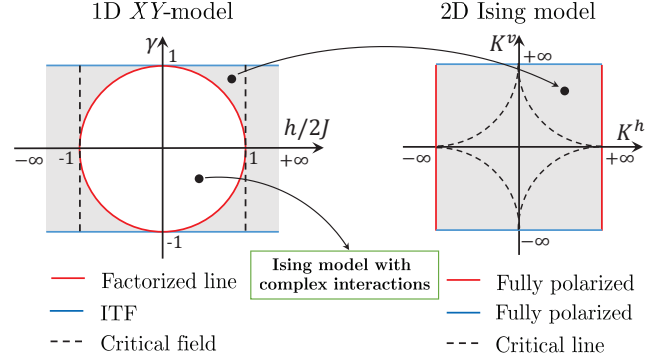}}
\caption{(Color online) The mapping of quantum $XY$-chain with uniform interactions onto a classical Ising model. Outside the circle is equivalent to a 2D square lattice Ising model with nearest neighbor interactions. Inside the circle may correspond to the Ising model with complex interactions. The unit circle, is the factorizing line which is mapped onto the Ising model with infinite horizontal coupling ($K^h\rightarrow \pm\infty$). The critical fields $h_c=\pm 2J$ correspond to critical lines, $\sinh 2K^v\sinh 2K^h=\pm 1$, in the Ising model. The blue lines represent the Ising model in a transverse field (ITF).}
\label{Equivalence}
\end{figure}

\section{Mapping onto 2D Ising model}\label{mapping}
By relating a transfer matrix associated with the classical system to the Hamiltonian of its corresponding quantum model, it has been proved by Suzuki \cite{suzuki1971relationship} that the two-dimensional Ising model is equivalent to the ground state of the quantum $XY$-chain with uniform interactions in the presence of a magnetic field, under appropriate relations among coupling parameters appearing in the two Hamiltonians.
His remarkable observation was that the Hamiltonian of $XY$-chain commutes with the transfer matrix of rectangular Ising
model for certain range of parameters. Two dimensional classical Ising model is expressed by the Hamiltonian
\begin{equation}\label{Ising}
\mathcal{H}_{\rm I}=-\sum_{i,j}(J^h_{j,j+1} \sigma^x_{ij} \sigma^x_{ij+1}+J^v_{i,i+1} \sigma^x_{ij} \sigma^x_{i+1j}),
\end{equation}
where $J^h$ and $J^v$ are horizontal and vertical couplings, respectively and $\sigma^{x}_{ij}=\pm 1$.
For the uniform couplings $J^h_{j,j+1}=J^{h}$ and $J^v_{i,i+1}=J^{v}$ the transfer matrix commutes with
the Hamiltonian of $XY$-chain ($\mathcal{H}_{q}$) with nearest-neighbor interactions, $J_{i,i+r}^{x,y}=J^{x,y}$, and uniform magnetic field $h_i=h$, if the coupling parameters are related via the following equations
\bea
&&  J^y/J^x=\tanh^2 K^{v*},\hspace{0.1 cm}\no \\
&&  h=2J^x\tanh K^{v*} \coth 2K^h,\label{uni-equ}
\eea
where $\ K^{v,h}= J^{v,h}/k_{\rm B} T$, $\tanh K^{v*}=\exp(-2K^{v})$ and $T$ is the temperature of the Ising system.
As a result of commutation of $\mathcal{H}_{q}$ and $V$, they can be diagonalized simultaneously and have a common set of eigenvectors. In particular the ground state of quantum Hamiltonian coincides with the eigenvector of transfer matrix with the maximum eigenvalue. This brings out several interesting results (\textit{i}) The critical temperature $T_c$ of the Ising model corresponds to the critical field of the $XY$ model, which is given by $h_c=2(J^x+J^y)$ (dashed lines in Fig. \ref{Equivalence}). Furthermore the region $T>T_c$ corresponds to the region $h>h_c$ (inside the diamond shape in Fig. \ref{Equivalence}) and $T<T_c$ to $h<h_c$ (outside the diamond shape in Fig. \ref{Equivalence}). (\textit{ii}) The spin correlations of the Ising model at temperature $T$ are related to those of $XY$ model in the ground state. (\textit{iii}) Scaling behavior and singularities of the two systems correspond to each other.

\begin{figure}[t]
\centerline{\includegraphics[width=8.5cm]{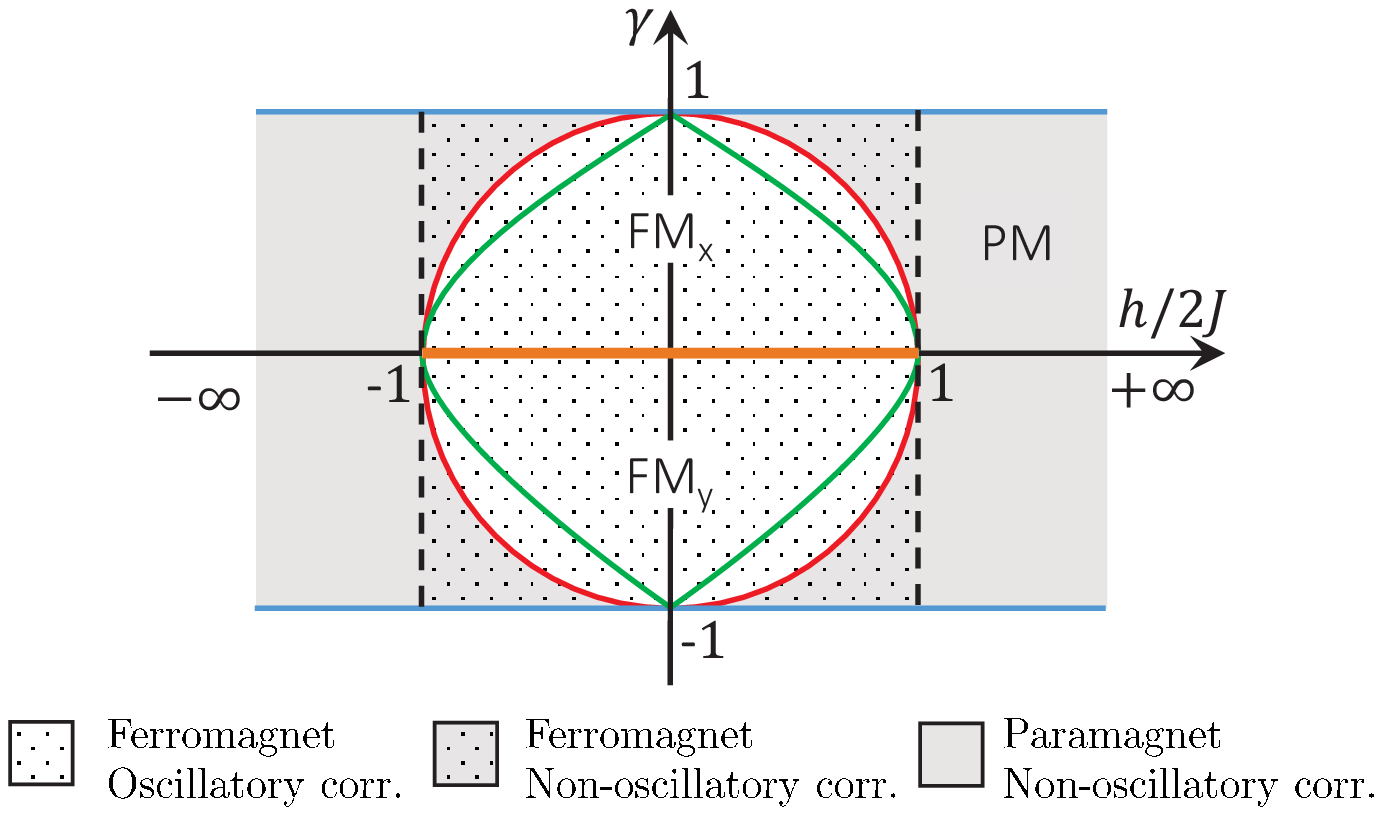}}
\caption{(Color online) Zero temperature phase diagram of quantum $XY$ chain in a transverse field. Factorizing circle separates the regions with oscillatory and non-oscillatory correlations. Dashed lines represent the transition from ferromagnetic (FM) to paramagnetic (PM) phase. Horizontal thick line shows the anisotropic transition from a ferromagnet with magnetization in $x$ direction to the one with magnetization in $y$ direction. Parabolic (green) lines, $\gamma^2\pm h/2J=1$, are the boundary between commensurate and incommensurate ferromagnetic phases.}
\label{XY}
\end{figure}

Equations (\ref{uni-equ}) have a noticeable property that do not have solutions for any given values of $J^x,J^y$ and $h$. As a result the region in which equivalence of the two systems holds is restricted to the domain where $h^2\geq 4J^xJ^y$. By taking $J^x=J(1+\gamma)$ and $J^y=J(1-\gamma)$,  where $|\gamma|\leq1$, this domain is outside the unit circle $(h/2J)^2+\gamma^2=1$ (see Fig. \ref{Equivalence}). The points on the circle are mapped to the Ising model with $K^h\rightarrow \pm\infty$ and $K^v=\text{finite}$. The Ising model with such anisotropic couplings, according to the exact result $M=[1-(\sinh 2K^h \sinh 2K^v)^{-2}]^{1/8}$, has maximum magnetization, $M=1$, which we have shown with fully polarized lines in Fig. \ref{Equivalence}.
Other properties of this circular boundary and its significance in the zero temperature phase diagram of the quantum XY-model are shown in Fig. \ref{XY}. The spin-spin correlations have non-oscillatory asymptotic behavior outside this unit circle
while they behave oscillatory inside it \cite{barouch1971statistical}.
Moreover, the circle is lying in the commensurate ferromagnetic phase which means the energy gap in the excitation spectrum always occurs at the Brillouin-zone boundary \cite{dennijs1988}.

It is also known that the ground state is factorized along this circle \cite{amico2009entanglement}.
Here, we show that this later property is a general feature of class of quantum $XY$-chains with arbitrary interactions. We first find the equivalence boundary of the generalized $XY$-chain with the rectangular Ising model and then show that this boundary coincides with the factorizing line of the quantum model.

\section{Boundary of equivalence and factorization}\label{Fac-Equ}
The above results, Eqs. (\ref{uni-equ}), have been generalized recently to the $XY$-chains with randomness and free boundary \cite{minami2014equivalence}. The conditions for commutation of the Hamiltonian of the $XY$-chain, Eq. (\ref{m.b.h}) with nearest neighbor interactions, and the transfer matrix of the Ising model;
\be
V=V_1^{1/2}V_2V_1^{1/2},
\label{TM}
\ee
where
\be
\no V_1=\Pi_i (e^{K^v_i}/\cosh K^{v*}_i) e^{\sum_i K^{v*}_i\sigma^z_i},~~
V_2 = e^{\sum_i K^{h}_i\sigma^x_i\sigma^x_{i+1}},
\ee
with $\ K^{v,h}_i=J^{v,h}_i/k_{\rm B} T$ and $\tanh K^{v*}_i=\exp(-2K^{v}_i)$,
are given in appendix (Eqs. (\ref{min-equivalence1}-\ref{min-equivalence5})). The solutions for these equations have been presented in Ref. \cite{minami2014equivalence} for special cases. Here we determine
the domain of existence of a solution for them.
For this purpose it is instructive to deduce the following equations by adding and subtracting Eqs. (\ref{min-equivalence1}) and (\ref{min-equivalence2}) or Eqs. (\ref{min-equivalence3}) and (\ref{min-equivalence4})
\be
h_i=\pm{\cal J}^{\pm}_{i-1,i} \coth (K^h_{i-1}\pm K^h_{i}),\label{couple}
\ee
where
\bea
{\mathcal J}^{\pm}_{i-1,i}=\sinh K_i^{v*} (\text{sech} K_{i+1}^{v*} J^x_i \pm \text{sech} K_{i-1}^{v*} J^x_{i-1}).
\label{calJ}
\eea
We have substituted for $J^y$ in terms of $J^x$ using the following relation
\be
\tanh K_i^{v*}  \tanh K_{i+1}^{v*} = J^y_i/J^x_i, \label{tahamtan}
\ee
obtained from Eq. (\ref{min-equivalence5}).
The first constraint imposed on parameters of $XY$-chain, as it is clearly seen from Eq. (\ref{tahamtan}) and positivity of $K_i^{v*}$, is that $J^x_i$ and $J^y_i$ should have the same sign and $|J^y_i/J^x_i|\leq 1$. For given exchange couplings $J^x_i$ and $J^y_i$  the vertical
couplings of the Ising model, $K_i^v$, will be fixed by Eq. (\ref{tahamtan}),
while the horizontal couplings, $K_i^h$, can be varied to find the allowed values of the magnetic field.
However, the couplings of the neighboring links should satisfy the constraint
\bea
\frac{\mathcal{J}^{+}_{i-1,i}}{\mathcal{J}^{-}_{i-1,i}}=-\frac{\coth(K^h_{i-1}-K^h_{i})}{\coth(K^h_{i-1}+K^h_{i})},
\label{constraint}
\eea
obtained by eliminating $h_i$ in Eqs. (\ref{couple}). The range of the hyperbolic functions
in Eqs. (\ref{couple}) and (\ref{constraint}),
place the following bounds on the values of the magnetic field at each site;
\bea \label{ineq}
 |h_i|  \geq
 \left\{\begin{array}{c}   |\mathcal{J}^{+}_{i-1,i}|, \ \text{if} \ \ |\mathcal{J}^{+}_{i-1,i}| \geq |\mathcal{J}^{-}_{i-1,i}| \\
                           |\mathcal{J}^{-}_{i-1,i}|, \ \text{if} \ \ |\mathcal{J}^{+}_{i-1,i}| \leq |\mathcal{J}^{-}_{i-1,i}|
 \end{array}\right..
\eea
The equality in Eq. (\ref{ineq}) gives the boundary of the equivalence, however a further care is needed on the signs of magnetic fields.
By applying the equation (\ref{constraint}) for two successive sites $i$ and $i+1$ of the $XY$-chain we find that there must be a
relation between the signs of magnetic fields $h_i$ and $h_{i+1}$ as $\sgn h_{i+1}=\sgn h_i$.
Finally the equation for the boundary reads as follows;
\be
h_i^b= \sinh K_{i+1}^{v*} (\text{sech} K_{i+1}^{v*} |J^x_i| +  \text{sech} K_{i-1}^{v*} |J^x_{i-1}|). \label{boundary}
\ee
In the special case of the uniform $XY$-chain $K_{i}^{v*}=K^{v*}$ and
the above equation simplifies to $h_i^b=2J\sqrt{1-\gamma^2}$ which is the circular boundary shown in Fig. \ref{Equivalence}.

Now we show that the corresponding coefficients in Eqs. (\ref{boundary}) and (\ref{ff}) and therefore $h_i^b$ and $h_i^f$ are identical. Since, $K^{v^*}_i>0$, $\sinh K^{v^*}_i\in(0,\infty)$, and $\text{sech} K^{v^*}_i\in(0,1)$ we can rewrite the hyperbolic functions in (\ref{boundary}) in terms of trigonometric functions as $\text{sech} K_i^{v*}=\sin \tilde{\alpha}_i$ and $\sinh K_i^{v*}=\cot \tilde{\alpha}_i$ with $\tilde{\alpha}_i\in(0,\pi/2)$. According to (\ref{tahamtan}) the angles $\tilde{\alpha}_i$ satisfy the equation $\cos \tilde{\alpha}_i  \cos \tilde{\alpha}_{i+1} = J^y_i/J^x_i$ which is the same as the equation of $\alpha_i$'s and therefore we conclude $\tilde{\alpha}_i=\alpha_i$. In other words, the equivalence boundary is the factorizing line of quantum $XY$-chain,
\be
h_i^b=h_i^f.
\ee
In the following we give more explicit results and solutions to the general equations within specific examples.

\subsection{Uniform factorizing field}
In addition to the uniform $XY$-chain, there are other spin models which also have a uniform factorizing field even though they have non-uniform exchange couplings.
Among them the spin-$1/2$ chains with ferro-antiferro ($f-a$) and antiferro-antiferro ($a-a$) bond alternations have been extensively investigated in the field of quantum magnetism and quantum information due to their rich field induced quantum phases such as Luttinger liquid phase \cite{sakai1995phase,abouie2008signature,mahdavifar2011effects}, dimerized phase and symmetry protected topological phases \cite{wang2013topological,canosa2010separability}.
The exchange interactions in these systems are
$J_{2i}^{\mu}=J_{a}^{\mu}<0$ and $J_{2i+1}^{\mu}=J_f^{\mu}>0$ for $a-f$ and $J_{2i}^{\mu}=J_{a}^{\mu}<0$ and $J_{2i+1}^{\mu}=\tilde{J}_a^{\mu}<0$
for $a-a$ models. For systems with $a-f$ bond alternations, the factorized ground state is $|FS\rangle=|\alpha,\alpha,-\alpha,-\alpha,\dots\rangle$ with $\cos\alpha=\sqrt{J^y_a/J^x_a}$ at the factorizing field,
\be
h^f=\sqrt{J^x_aJ^y_a}(1+|J_f^x/J^x_a|).
\label{hf-dimer1}
\ee
In systems with $a-a$ bond alternations, the factorized ground state is the Neel state, $|FS\rangle=|\alpha,-\alpha,\alpha,-\alpha,\dots\rangle$, with $\cos\alpha=\sqrt{J^y_a/J^x_a}$ and the factorizing field is:
\be
h^f=\sqrt{J^x_aJ^y_a}(1+\tilde{J}_a^x/J^x_a).
\label{hf-dimer2}
\ee
The bond alternating quantum spin chains are mapped to a 2D Ising model with uniform vertical interaction, $K_j^v=K^v$, and horizontal $a-f$ bond alternations, $K_{2i}^h=K_a^h$ and $K_{2i+1}^h=K_f^h$, or $a-a$ bond alternations, $K_{2i}^h=K_a^h$ and $K_{2i+1}^h=\tilde{K}_a^h$.
In these cases the equivalence boundary (\ref{boundary}) is simplified to
\be
h^b=\tanh K^{v*}|J_a^x|(1+|J_c^x/J_a^x|),
\ee
where $\tanh K^{v*}=\sqrt{J^y_a/J_a^x}$ and $J_c^x=J_f^x$ or $\tilde{J}^x_a$. These boundaries exactly coincide with the factorizing lines, (\ref{hf-dimer1}) and (\ref{hf-dimer2}) of XY-chains with bond alternations.

\subsection{Non-uniform factorizing field}
In many cases, we can not get a factorized ground state by applying a uniform magnetic field.
Some of them, for example trimerized and tetramerized spin chains\cite{PhysRevB.73.134427,PhysRevB.78.134415}, have a factorized ground state in the presence of a non-uniform field which is sum of a uniform and a staggered field \cite{abouie2012ground}.
In a tetramerized spin chain with the exchange couplings  $J^{\mu}_{4i}=J^{\mu}_{4i+1}=J^{\mu}_{a}<0$ and $J^{\mu}_{4i+2}=J^{\mu}_{4i+3}=J^{\mu}_{f}>0$, the factorizing fields are given by $h_{4i}^f=h_{4i+2}^f=h^f_u$, $h_{4i+1}^f=h^f_u+h^f_s$, and $h_{4i+3}^f=h^f_u-h^f_s$, where
\bea
h^f_u&=&\sqrt{J_a^xJ_a^y}(1+|J_f^x/J_a^x|),\no \\
h^f_s&=&\sqrt{J_a^xJ_a^y}(1-|J_f^x/J_a^x|).\label{huhs}
\eea
The corresponding factorized ground state in this case is $|\alpha,\alpha,\alpha,-\alpha,\dots\rangle$, where $\cos\alpha=\sqrt{J_a^y/J_a^x}$.
The equivalent classical model is a 2D tetramerized Ising model with $K^h_{4i}=K^h_{4i+1}=K_a^h$, $K^h_{4i+2}=K^h_{4i+3}=K_f^h$ and $K_i^v=K^v$.
Using Eq. (\ref{boundary}) we find that
$h_{4i}^b=h_{4i+2}^b=\tanh K^{v^*} (|J_a^x|+J_f^x)$, $h_{4i+1}^b=2\tanh K^{v^*} |J_a^x|$, and $h_{4i+3}^b=2\tanh K^{v^*} |J_f^x|$. By substituting  from Eq. (\ref{tahamtan}) for $\tanh K^{v^*}=\sqrt{J_a^y/J_a^x}$ and using Eqs. (\ref{huhs}) one can see that these boundary fields are exactly the factorizing fields of the tetramerized spin chain.

\subsection{Non-factorizable models}
There are models in which it is not possible to factorize the ground state by applying any values of magnetic fields. For example, the model with exchange interactions $J^x_i=1-(-1)^i\delta$ and $J^y_i=1+(-1)^i\delta$, where $0<\delta<1$ does not possess a factorizing field because $|J^x|< |J^y|$ at even links while $|J^x| > |J^y|$ for odd links. This property does not permit a real solution for angles $\theta_i$ in Eq. (\ref{coscos}). On the other hand, Eq. (\ref{tahamtan}) imposes a similar limitation in mapping this model onto the classical Ising model.
This means that there is a relationship between the non-factorizability of the ground state and the impossibility of the mapping onto the Ising model. It should be mentioned that the model with $|J^x| > |J^y|$ in all links could be transformed to the one with $|J^x| < |J^y|$ by a global rotation of $\pi/2$ around $z$ axis therefore does have a factorizing field. This transformation is not possible even if there is only one link with $|J^x| > |J^y|$ while all other links have $|J^x| < |J^y|$. In conclusion, the non-factorizability of the ground state of quantum model implies the impossibility of the mapping onto the classical Ising model and therefore the absence of the equivalence boundary for such models.

\section{Summary and conclusion}\label{sum}
In summary, we have provided a connection between ground state factorization of quantum $XY$-chains with arbitrary interactions and their mappability onto a two dimensional classical Ising model. By making use of Suzuki's criterion for equivalence, i.e. the commutativity of $\mathcal{H}_q$ and $V$, we obtained the natural boundary for the general $XY$-model and showed that it is identical to the factorizing line in parameter space.  This allows us to discuss the counterparts of the factorization and vanishing the entanglement of the ground state in the behavior of equivalent classical model. We have applied our results to several examples and shown that in the cases with a factorizing field, this line coincides with the boundary obtained from mapping to classical model while the cases with no factorizing field do not have equivalent Ising model. Here we have considered the infinite chains however the results are also applicable to finite chains with different boundary conditions. The presented correspondence suggests a promising approach for obtaining the factorizing field of quantum systems. Unlike the standard approach \cite{kurmann1982antiferromagnetic} this method is not restricted to the systems with two-particle interactions and frustration. Another benefit of this approach for the future work is that it might lead to a criterion for factorizability of the ground state of the Hamiltonians based on their symmetries, since the commutation with the transfer matrix indicates some hidden symmetry of the Hamiltonian. The generalization of the results to systems with $XYZ$ interactions, frustration and higher dimensions will be investigated in the future work.

\acknowledgments
RS would like to acknowledge the financial support of University of Tehran for this research under grant number 28957/01/1.

\section{Appendix: Mapping conditions}

The conditions for commutation of the Hamiltonian of the $XY$-chain, Eq. (\ref{m.b.h}) with nearest neighbor interactions, and the transfer matrix of the Ising model, Eq. (\ref{TM}),
are given by \cite{minami2014equivalence}
{\medmuskip=0mu
\thinmuskip=0mu
\thickmuskip=0mu
{\begin{flalign}
& h_i \tilde{p}^+_i-2[J^-_{i-1}(K^h_{i-1}\tilde{q}^{+}_i + K^h_i \tilde{q}^{-}_i)+J^+_i(K^h_{i-1}\tilde{q}^{-}_i +K^h_i \tilde{q}^{+}_i)] = h_i,\label{min-equivalence1}  \\
& h_i \tilde{p}^-_i-2[J^-_{i-1}(K^h_{i-1}\tilde{q}^{-}_i + K^h_i \tilde{q}^{+}_i)+J^+_i(K^h_{i-1}\tilde{q}^{+}_i + K^h_i \tilde{q}^{-}_i)] =0,\label{min-equivalence2}   \\
& J^-_{i-1}(\tilde{p}^+_i+1)-2h_i (K^h_{i-1}\tilde{q}^{+}_i + K^h_i \tilde{q}^{-}_i)+J^+_i\tilde{p}^-_i = 0,\label{min-equivalence3}  \\
& J^+_i(\tilde{p}^+_i+1)-2h_i (K^h_{i-1}\tilde{q}^{-}_i + K^h_i \tilde{q}^{+}_i)+J^-_{i-1}\tilde{p}^-_i = 0,\label{min-equivalence4} \\
& p^-_i J^x_i-p^+_i J^y_i=0,\label{min-equivalence5}
\end{flalign}}}
where
{\medmuskip=0.1mu
\thinmuskip=0.1mu
\thickmuskip=0.1mu
\begin{eqnarray}
p^{\pm}_i &=&  (\cosh (K^{v*}_{i}+K^{v*}_{i+1})\pm\cosh (K^{v*}_{i}-K^{v*}_{i+1}))/2,\no   \\
q^{\pm}_i &=& (\sinh (K^{v*}_{i}+K^{v*}_{i+1})\pm\sinh (K^{v*}_{i}-K^{v*}_{i+1}))/2,\no   \\
\tilde{p}^{\pm}_i &=&  (\cosh 2(K^h_{i-1}+K^h_i)\pm\cosh 2(K^h_{i-1}-K^h_i))/2, \no  \\
\tilde{q}^{\pm}_i &=&  (\text{sinhc} 2(K^h_{i-1}+K^h_i)\pm\text{sinhc} 2(K^h_{i-1}-K^h_i))/2, \no  \\
J^{\pm}_i &=& q^{\pm}_i J^x_i - q^{\mp}_i J^y_i \no,
\end{eqnarray}}
and $\text{sinhc}\hspace{0.07 cm}x=\sinh x/x$.

\bibliographystyle{eplbib}
\bibliography{QC-refs}

\begin{thebibliography}{10}
\expandafter\ifx\csname url\endcsname\relax\def\url#1{\texttt{#1}}\fi

\bibitem{kurmann1982antiferromagnetic}
\Name{Kurmann J., Thomas H. \and M{\"u}ller G.} \REVIEW{Physica A: Statistical
  Mechanics and its Applications}{112}{1982}{235}.

\bibitem{giampaolo2008theory}
\Name{Giampaolo S.~M., Adesso G. \and Illuminati F.} \REVIEW{Physical review
  letters}{100}{2008}{197201}.

\bibitem{giampaolo2009separability}
\Name{Giampaolo S.~M., Adesso G. \and Illuminati F.} \REVIEW{Physical Review
  B}{79}{2009}{224434}.

\bibitem{rossignoli2009factorization}
\Name{Rossignoli R., Canosa N. \and Matera J.} \REVIEW{Physical Review
  A}{80}{2009}{062325}.

\bibitem{rezai2010factorized}
\Name{Rezai M., Langari A. \and Abouie J.} \REVIEW{Physical Review
  B}{81}{2010}{060401}.

\bibitem{abouie2012ground}
\Name{Abouie J., Rezai M. \and Langari A.} \REVIEW{Progress of Theoretical
  Physics}{127}{2012}{315}.

\bibitem{amico2009entanglement}
\Name{Amico L. \and Fazio R.} \REVIEW{Journal of Physics A: Mathematical and
  Theoretical}{42}{2009}{504001}.

\bibitem{roscilde2004studying}
\Name{Roscilde T., Verrucchi P., Fubini A., Haas S. \and Tognetti V.}
  \REVIEW{Physical review letters}{93}{2004}{167203}.

\bibitem{ciliberti2010quantum}
\Name{Ciliberti L., Rossignoli R. \and Canosa N.} \REVIEW{Physical Review
  A}{82}{2010}{042316}.

\bibitem{tomasello2011ground}
\Name{Tomasello B., Rossini D., Hamma A. \and Amico L.} \REVIEW{EPL
  (Europhysics Letters)}{96}{2011}{27002}.

\bibitem{campbell2013criticality}
\Name{Campbell S., Richens J., Gullo N.~L. \and Busch T.} \REVIEW{Physical
  Review A}{88}{2013}{062305}.

\bibitem{sarandy2013quantum}
\Name{Sarandy M.~S., De~Oliveira T.~R. \and Amico L.} \REVIEW{International
  Journal of Modern Physics B}{27}{2013}{1345030}.

\bibitem{rossignoli2008entanglement}
\Name{Rossignoli R., Canosa N. \and Matera J.} \REVIEW{Physical Review
  A}{77}{2008}{052322}.

\bibitem{canosa2010separability}
\Name{Canosa N., Rossignoli R. \and Matera J.~M.} \REVIEW{Physical Review
  B}{81}{2010}{054415}.

\bibitem{giorgi2009ground}
\Name{Giorgi G.~L.} \REVIEW{Physical Review B}{79}{2009}{060405}.

\bibitem{suzuki1971relationship}
\Name{Suzuki M.} \REVIEW{Progress of Theoretical Physics}{46}{1971}{1337}.

\bibitem{barouch1971statistical}
\Name{Barouch E. \and McCoy B.~M.} \REVIEW{Physical Review A}{3}{1971}{786}.

\bibitem{dennijs1988}
\Name{den Nijs M.} \REVIEW{\textit{Phase Transitions and Critical Phenomena}
  Edited by C. Domb and J. L. Lebowitz, vol. 12, page 264}{}{1988}{}.

\bibitem{minami2014equivalence}
\Name{Minami K.} \REVIEW{EPL (Europhysics Letters)}{108}{2014}{30001}.

\bibitem{sakai1995phase}
\Name{Sakai T.} \REVIEW{Journal of the Physical Society of
  Japan}{64}{1995}{251}.

\bibitem{abouie2008signature}
\Name{Abouie J. \and Mahdavifar S.} \REVIEW{Physical Review
  B}{78}{2008}{184437}.

\bibitem{mahdavifar2011effects}
\Name{Mahdavifar S. \and Abouie J.} \REVIEW{Journal of Physics: Condensed
  Matter}{23}{2011}{246002}.

\bibitem{wang2013topological}
\Name{Wang H.~T., Li B. \and Cho S.~Y.} \REVIEW{Physical Review
  B}{87}{2013}{054402}.

\bibitem{PhysRevB.73.134427}
\Name{Gu B., Su G. \and Gao S.} \REVIEW{Phys. Rev. B}{73}{2006}{134427}.

\bibitem{PhysRevB.78.134415}
\Name{Rachel S. \and Greiter M.} \REVIEW{Phys. Rev. B}{78}{2008}{134415}.

\end{thebibliography}

\end{document}